# A Conceptual Network-Based Approach to Inferring the Cultural Evolutionary History of the Baltic Psaltery

**Tomas Veloz (tomas.veloz@ubc.ca)**
Department of Mathematics, University of British Columbia
3333 University Way, Kelowna BC, V1V 1V7, Canada

**Ilya Tëmkin (itemkin@nvcc.edu)**
Biology Department, Northern Virginia Community College, Annandale, VA, 22003
and Department of Invertebrate Zoology, National Museum of Natural History, Washington, DC 20013, USA

**Liane Gabora (liane.gabora@ubc.ca)**
Department of Psychology, University of British Columbia
3333 University Way, Kelowna BC, V1V 1V7, Canada

**Abstract**

The application of conventional phylogenetic techniques for inferring cultural history is problematic due to differences in the nature of information transmission in biological and cultural realms. In culture, units of transmission are not just measurable attributes, but communicable concepts. Therefore, relatedness amongst cultural elements often resides at the conceptual level not captured by traditional phylogenetic methods. This paper takes a cognitively inspired approach to analyzing material cultural history. We show that combining data for physical attributes of cultural artifacts with conceptual information can uncover cultural influences among different ethnolinguistic groups, and reveal new patterns of cultural ancestry. Using the Baltic psaltery, a musical instrument with a well-documented ethnographic and archaeological record, we recovered a previously unacknowledged pattern of historical relationship that is more congruent with geographical distribution and temporal data than is obtained with other approaches.

## Introduction

The artifacts we put into the world reveal much about the minds that conceived them. The evolutionary history of human artifacts tells the story of how our thoughts, beliefs, and understanding of the world we live in, has unfolded over the ages. Using tools and techniques that include insights from cognitive science, we are starting to piece this exciting story together.

Phylogenetic approaches to reconstructing evolutionary patterns and processes, applied routinely in systematics, are increasingly applied not just to linguistics, but also elements of material culture, such as textiles, weapons, and musical instruments (*e.g.*, Collard, Shennan & Tehrani, 2006; Forster & Toth, 2003; Mace & Holden, 2005; Shennan, 2008; Whiten *et al.,* 2011). Originally developed in biology for inferring historical relationships among groups of organisms, phylogenetics makes use of assumptions about how information is organized and transmitted that reflects peculiarities of the biology world.

The direct transfer of methodology from biology to culture has raised the question about the extent to which meaningful parallels can be drawn between the processes of change in the two domains (*e.g.*, Eldredge, 2000; Gabora, 2006). Application of phylogenetics to material culture assume that the same (or analogous) causal processes operate in culture and nature. However, what is transmitted through culture is not just the objects themselves, but rather communicable perspectives and concepts, such as notions of complementarity (*e.g.*, between a mortar and pestle, which share no attributes but clearly are related), or competition for the same cultural niche (e.g., spear, gun, rope, and so forth) that may or may not be reflected in the artifact design. Indeed, some claim that the differences between biological and cultural evolution are so insurmountable that insights obtained from biology are completely irrelevant in a cultural context (Moore, 1994; Dewar, 1995; Terrell, 2001). Others such as ourselves take a more moderate stance, arguing that well there are significant parallels as well as differences between biological and cultural systems, phylogenetic techniques have limited application to culture, and it is necessary to either significant modify existing approaches, or develop altogether new ones (Eerkens, Bettinger, & McElreath, 2005; Borgerhoff-Mulder *et al*., 2006; Gabora, 1998, 2006, 2008; Nunn *et al*., 2006; Tëmkin & Eldredge, 2007).

In a previous paper (Gabora *et al.,* 2011) we put forward a graph theory-based approach to modelling the evolution of cultural artifacts, and applied it to a well-studied set of artifacts: early projectile points from the Southeastern United States. This data set had previously been modelled using a phylogenetic approach (O'Brien, Darwent, & Lyman, 2001), and using an earlier version of the network-based approach, upon which our model is based (Lipo, 2005). The model included reticulate relationships as well as hierarchical groupings, and incorporated conceptual information to complement physical attribute data. We showed that incorporating conceptual information that is not typically captured by the phylogenetic analysis can significantly alter the inferred pattern of historical relationships amongst artifacts.

The current paper reports on new developments of the model, most notably, a means of evaluating the relative contributions of different types of data to historical inference. In addition, we apply the approach to a very different domain, thus demonstrating its generalizability.

## The Data Set

The experimental data set is a representative selection of Baltic psalteries, a traditional plucked stringed musical instrument distributed among Baltic, Finnic, and Slavic peoples of Northeastern Europe. Until recently, the Baltic psaltery remained an integral part of secular and ritual life, and has become a national symbol for every ethnic group that has it. The origin and historical development of the Baltic psaltery has been a controversial subject for over a century (reviewed by Raynolds, 1984) and remains so to this day (Povetkin, 1989; Haas, 2001; Tëmkin, 2004).

The data on psalteries consist of extensive descriptions of structural and ornamental features, and documented (or inferred) playing styles for 13 ethnographic (dated by 17-20 centuries) and two archaeological (dated by late 10-13 centuries) artifacts, representing major pertinent ethnolinguistic groups. The data set includes two Estonian (EST), two Finnish (FIN), three Latvian (LAT), three Lithuanian (LIT), three Russian (RUS), and two presumably Slavic archaeological instruments from Novgorod, northwestern Russia (NVG).

## The Conceptual Network Approach

In this section we outline our approach. We begin by summarizing how it models attributes and concepts. We then move on to the conceptually new contribution of this paper, the use of 'perspectives' to bias the network structure in culturally meaningful ways.

### The Structure of Attributes and Concepts

Following convention, concepts are indicated by all capital letters (PSALTERY), whereas an actual artifact, or instance of a psaltery is indicated with all small letters (psaltery). The more superficial level of conceptual structure consists of what Rosch (1978) refers to as *basic level concept,* such as PSALTERY, which mirror classes of objects that share a broad range of perceivable attributes. These basic level concepts can be recursively differentiated. For example, a psaltery's attribute "strings" can be differentiated into "metal" or "nylon" depending on the type of the material the strings are made of. Each of these subordinate attributes can be further resolved by introducing their respective attributes, such as, for instance, "metal type," "nylon type," or "color." Some attributes of the second degree may be shared by those of the first: both metal and nylon strings have color, but only metal strings are made from a specific metal type. Hence, a basic level concept can be represented as a root of a graph and its attributes arranged by levels of descriptive resolution. Because attributes at a given level can be connected to multiple attributes of levels above and below, the resultant structure contains both hierarchical and reticulate aspects.

Basic level concepts are generalized at a more abstract level as instances of superordinate concepts, such as MUSICAL INSTRUMENT. Superordinate concepts typically refer to multiple basic level categories (*e.g.*, MUSICAL INSTRUMENT consists of both PSALTERY and CORNET).

### Conceptual Structure and its Representation

Each artifact is represented by a network of attributes consisting of reticulated hierarchies. The attributes can be physical or non-physical (conceptual). The total network of all available attributes constitutes the conceptual structure (Figure 1). Each artifact is represented by a subnetwork of the conceptual structure. (In a sense this is conceptually similar to phylogenetic approaches where all taxa and their attributes are described as arrays of specific character states in a character state data matrix.)

The conceptual structure introduced here is based on the state context property (SCOP) theory of concepts (Aerts & Gabora, 2005a, 2005b; Gabora & Aerts, 2002) and is equivalent to a simplified ontology (Sowa, 2000). We incorporate the notion of context by introducing the notion of *perspective* (see next section), and use graph theory

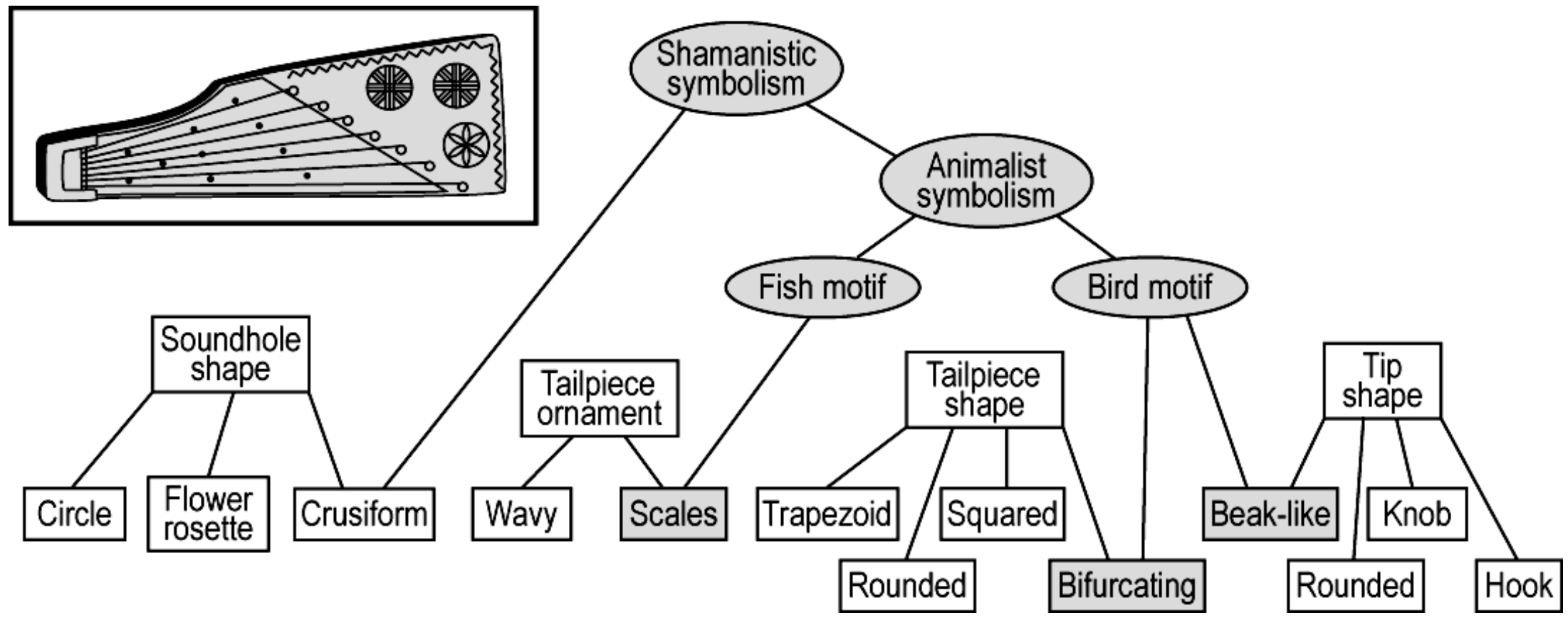

Figure 1: A segment of the conceptual structure used in the Baltic psaltery analysis. Elements in square boxes represent

instead of operator theory to develop similarity metrics.

## Incorporation of Perspectives

Given the heterogeneity of attributes, (structural *vs*. ornamental, physical *vs*. conceptual, and so forth), it may be useful to be able to explore the effect of different sets of attributes on the evolutionary pattern. Depending on the perspective from which a given artifact is considered, the emphasis is placed on a particular set of attributes. For example, a musical instrument can be viewed as an art object, a product of craftsmanship, a sound-producing device, or a sacred symbol. Given a particular perspective, a set of attributes may include physical features (*e.g.*, presence of a soundhole), conceptual non-physical descriptors (*e.g.*, ritualistic function), or a mixture of both. Because physical and non-physical attributes represent different culturally relevant aspects of the artifact, including both types of data in the analysis can potentially produce a more comprehensive, and presumably, more realistic pattern of evolutionary history for a collection of artifacts as a whole.

This is accomplished by defining a *perspective,* a part of the conceptual structure that includes a predefined subset of attributes that may or may not be directly linked to each other. Multiple perspectives can be defined for a given conceptual structure, and they may or may not have attributes in common. In the analysis of the Baltic psaltery, we defined three perspectives: (1) *Physical Attributes*, containing structural or decorative features of actual artifacts; (2) *Performance,* containing concepts related to music performance styles; and (3) *Symbolism*, containing concepts or physical attributes associated with sacred symbolism and ritualistic significance. Perspectives in the conceptual network approach are, in a sense, analogous to character partitions in biological phylogenetics.

## Reliablity and Similarity

In the proposed approach, similarity among artifacts is assessed by pairwise comparison of their network representations. The comparisons can be made between entire sub-networks corresponding to complete representations of the two artifacts, or with respect to a particular perspective, or set of perspectives. To formalize this notion of similarity, we assume two artifacts, $a$ and $a'$, and a perspective $p$. The attributes associated with $p$ are designated $V(p)$, and the relevant attributes of $a$ and $a'$ with respect to $p$ are designated $V(a)$ and $V(a')$. We introduce two functions:

$$O(a,a',p) = s(V(a) \cap V(a') \cap V(p)) \qquad (1)$$

$$D(a,a',p) = s(V(a) \cap V(p)) + s(V(a') \cap V(p)) - 2O(a,a',p) \qquad (2)$$

where $s$ is a measure of attribute similarity, $O(a,a',p)$ is the *overlap* and $D(a,a',p)$ is the *divergence* between artifacts $a$ and $a'$ with respect to $p$. The overlap and divergence account for the number of attributes included in $p$ that are shared or non-shared, respectively, by $a$ and $a'$.

Accounting for both the overlap and divergence is critical to determine the similarity of two artifacts. Overlap alone can lead to an overestimation of overall similarity in some situations, such as in a trivial case where one of two artifacts possesses just one attribute included in a particular perspective and this attribute is shared by another artifact that has a greater number of attributes included in the same perspective. In this case, the failure to account for divergence will erroneously interpret the complete overlap as absolute similarity between the artifacts with respect to the chosen perspective.

Some perspectives capture more of the overlap and divergence between artifacts than others. Indeed, a (non empty) perspective may include none of the attributes of the artifacts being compared, or it may include them all. In general, some portion of the total overlap and divergence between two artifacts is captured by a perspective. We define the *reliability* $R(p,a,a')$ as the proportion of the overlap and divergence between artifacts $a$ and $a'$ given perspective $p$ as follows:

$$R(p,a,a') = (O(a,a,p) + O(a',a',p)) / (|a|+|a'|) \qquad (3)$$

where $|a|$ and $|a'|$ are the size of the graphical representations of $a$ and $a'$ respectively. Note that the entire conceptual structure can also be considered as a perspective. In this case, its reliability is equal to 1 for any given pair of artifacts. This is because it contains all possible concepts used to represent each artifact. Hence, the whole conceptual structure, and more inclusive perspectives in general, have greater reliability. However, the notion of reliability in itself may not be sufficient as an estimation of the perspective's effect on the similarity between artifacts. There may exist a small portion of the conceptual structure such that if considered as a perspective it would have small reliability, but which may nevertheless be vital for establishing the similarity of some artifacts. Therefore, given a set of perspectives $P = \{p_1,\ldots, p_n\}$, we introduce a *perspective weight vector,* $\{v_1,\ldots, v_n\}$, which gives the relative degree of importance of $p$ in $P$, and defines the *similarity* between two artifacts $a$ and $a'$ with respect to $P$ by the following formula:

(4)

The *similarity* $S(P,V,a,a')$ takes into account the overlap and divergence between their graph representations summed over all perspectives, each weighted by their respective reliability values and perspective weight vectors. Note that the greater the reliabilities of the perspectives in $P$, the greater the similarity between $a$ and $a'$.

The approach provides a means of exploring the effect of individual perspectives on evolutionary inference. There are several ways of going about this, ranging in objectivity from fully automated, unbiased naive models to sophisticated expert-specified models that allow for incorporation of background information: (1) a *uniform weights* model, which weighs all perspectives equally; (2)

an *implied weighting* model, which weighs each perspective proportional to its reliability; (3) a *sensitivity* model, which explores a range of weights allowing for identification of most and least stable relationships; and (4) an *expert choice* model, which enables the user to specify unique weights (including removing a selected perspective from the analysis) to each perspective.

## Similarity Graph and Cultural History

A pairwise similarity matrix based on comparing all pairs of artifacts with respect to a chosen perspective (or set of perspectives), is used to compute a *similarity graph* where each vertex (node) corresponds to an artifact, and an edge implies that the extreme vertices of the edge are similar. The edge is labelled by the similarity weight between the two artifacts. The graphical representation of the similarity graph can subsequently be interpreted as a historical pattern of relationships amongst included artifacts. It is important to emphasize that the similarity graph is *not* a cultural phylogeny, but a mere representation of similarity among artifacts: it does not incorporate explicit actual cultural transmission models but provides an independent framework for establishing historical hypotheses that can corroborate or disagree with existing models of cultural change. We restrict ourselves to the *maximal similarity graph,* which connects each artifact to only those that have the highest similarity to it (in most cases, for each artifact there is only one most similar artifact). The maximal similarity graph provides an approximation to the artifact's true cultural history.

## Computational Implementation

The program we use to infer cultural lineages was developed using the object-oriented Java platform with extension packages for working with networks (JUNG). It allows for the creation of a conceptual structure by adding nodes (concepts) and edges (conceptual relationships). Perspectives and artifacts can be generated as well. One can specify the entries of the perspective weight vector using an array of sliders (one slider is automatically created for each perspective the user creates). Other software functions allow the user to export and import these structures for later use. The currently implemented default weighting scheme is the *implied weighting* model, which weighs each perspective proportional to its reliability. By modifying the perspective weights, the user can recompute the similaries, and visualize the resulting changes to the similarity graph. This enables exploration of the resulting similarity graphs found in different regions of perspective weight space.

# Results

We present the patterns of relationship obtained for the Baltic psalteries with respect to two perspectives: *Physical Attributes* and *Symbolism*. When only physical attributes of the psalteries are considered, the resulting similarity graph recovers clusters of instruments corresponding to ethnolinguistic affinities with the exception of the Baltic instruments that appear to be more dispersed (forming three lineages; Figure 2A). This is consistent with previous results based on maximum parsimony analysis (Tëmkin, 2004; Tëmkin and Eldredge, 2007). On the other hand, when only symbolic aspects are taken into account, such sharp delineation based on the linguistic affinity becomes less evident and novel relationships emerge (Figure 2B). For example, in the former scenario, the Baltic instruments were linked to the Estonian (Finnic) ethnographic instruments, whereas in the latter they have no connection with the Estonian instruments, and display a novel connection with the Slavic instruments.

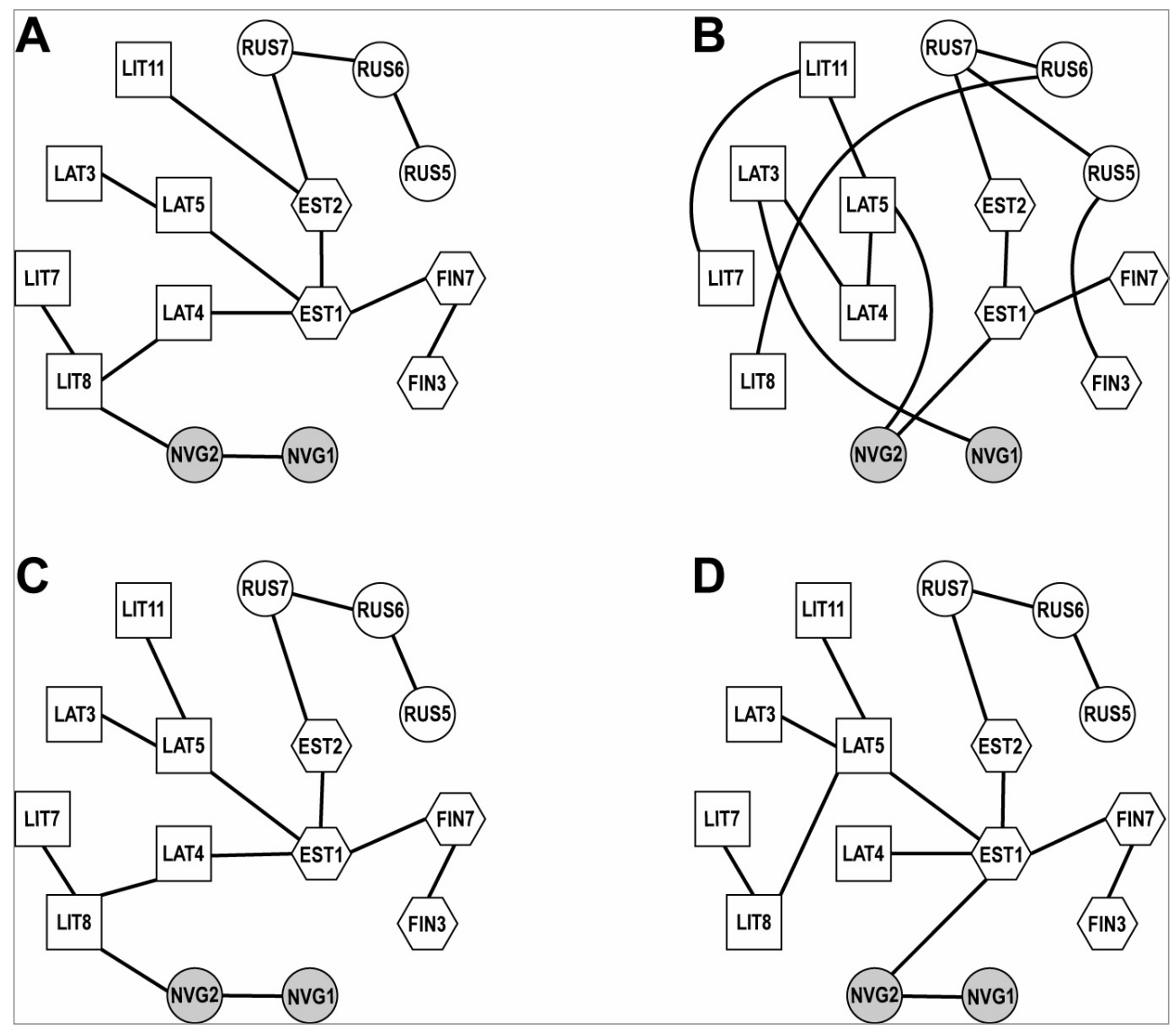

Figure 2. Similarity graphs based on the conceptual network analysis of Baltic psalteries under different perspective weighting schemes. (A) *Physical attributes*; (B) *Symbolism*; (C) *Physical attributes* and *Symbolism* (equal weights); (D) *Physical attributes* (25% weight) and *Symbolism* (75% weight). Each node corresponds to a single artifact. The node shapes indicate ethnolinguistic groups: Slavic (circle), Finnic (hexagon), and Baltic (square). Shaded nodes designate archaeological instruments (10-13 cc.); the remaining nodes correspond to ethnographical instruments (17-20 cc).

Such incongruence in similarity inference between the two perspectives suggests that the constructive principles of the psalteries were regionally constrained and possibly insured by master and apprentice relationship which has a strong linguistic component. Similarity in symbolic elements across cultures, however, appears to correspond more closely to the geographic proximity and, possibly, some symbolic features spread as decorative designs without affecting structural aspects of local musical instrument making traditions.

When both perspectives are analyzed simultaneously under equal weights, the resulting similarity graph is similar to those based on the analysis of physical attributes alone, but results in greater similarity amongst the Baltic instruments (Figure 2C). This congruence between the two graphs can be accounted for by a much greater number of physical (80) than symbolic (20) attributes.

To reveal the impact of symbolic attributes given the situation that they were outnumbered by physical attributes, the data set was re-analyzed with 25% and 75% weights for the *Physical Attributes* and *Symbolism* perspectives respectively. The resulting similarity graph was identical to the results of the analysis under equal-weight with respect to the relationships among the Baltic, Finnic, and Slavic ethnographical instruments with two significant differences (Figure 2D). First, the relationships among the Baltic instruments were stronger (as five out of 6 instruments formed a single cluster). Second, the connection of most ancient, archaeological instruments shifted from the Baltic to the Finnic instruments.

The most comprehensive cladistic analysis of the Baltic psaltery failed to unequivocally resolve which of the three groups of instruments (Baltic, Finnic, or Slavic) were more closely related to the root of the tree, the medieval artifacts from Novgorod in northwestern Russia (Tëmkin & Eldredge, 2007). The recovery of the Lithuanian (Baltic) psaltery as most basal agrees with a presumed northward diffusion of the instrument (the second wave of dispersal) in (Tëmkin, 2004). The alternative scenario, in which the Novgorodian instruments bear greater similarity to the Finnic instruments (largely attested by shared symbolic significance) suggests an intriguing hypothesis for interpreting the instrument's history as it is more consistent with archaeological data which indicates that medieval Novgorod, where most ancient Baltic psalteries were discovered, had a substantial proportion of Finnic population (Tõnurist, 1977).

## Discussion

By fusing physical information about artifacts with the conceptual information our ancestors were using to *create* these artifacts, we arrive at a more accurate picture of the evolutionary trajectories by which the artifacts evolved, and by which our human understanding of the world took shape. This paper develops the conceptual and mathematical foundation for a novel approach to reconstructing patterns of cultural evolutionary history based on graph theory. It uses algorithms for constructing and displaying similarity graphs that can be biased by conceptual knowledge from different domains. This approach circumvents the limitations of traditional phylogenetic approaches by (1) allowing for simultaneous analysis of cognitive information and physical character data, (2) providing the means for evaluating relative contributions of different types of data to historical inference, and (3) expanding hierarchical approach to include reticulate relationships.

We tested the utility of the conceptual network approach for inferring historical patterns of relatedness amongst artifacts by applying it the analysis of the Baltic psaltery, a stringed musical instrument unique to northwestern Europe. Not only did the approach capture the essential features of the instrument's history inferred previously using other methods, it also provided new insights that invite a novel interpretation of the instrument's evolution. To achieve this, it was necessary to distinguish between hierarchically organized conceptual attributes (largely pertaining to sacred symbolic imagery), and physical characteristics (such as elements of the instrument's construction). Although this was readily accomplished with the current approach, it cannot be done with other approaches. The approach is not limited to a specific data types; it can be extended to include linguistic information, or any other discrete character information.

## Future Directions

Although limited in its present formulation, the conceptual network approach provides large number of avenues for future theoretical and mathematical developments. We are currently developing methods for computing the expected reliability of perspectives and determining the range of weights (parameter space) in which a given perspective becomes significant in the similarity computation. Other immediate plans include developing sophisticated software for analysis of the conceptual network.

In this investigation, we only explored the maximal similarity graph, which connects each artifact to the artifact that has the highest similarity to it according to the chosen perspective weight vector (first neighbours graph). In future investigations, we will consider the second or further most similar artifacts ($n^{th}$ neighbour similarity graph), and establish a mechanism to automatically split the perspective space according to different criteria of similarity graph equivalence. This will enable us to explore the conglomeration of similarity graphs obtained from the different sets of perspective weight vectors. To further refine the approach, we will construct similarity graphs not only considering the most similar connections of each artifact, but connections above a certain threshold. This could reveal different similarity ranges where the parameter space is split in qualitatively different ways.

Finally, we plan to investigate the applicability of Kemp and Tenenbaum (2008) structure discovery models.

## Acknowledgments

We would like to acknowledge grants to Liane Gabora from the *Social Sciences and Humanities Research Council of Canada,* the *Natural Sciences and Engineering Research Council of Canada,* and the Concerted Research Program of the *Fund for Scientific Research of Belgium*.